\documentclass{PoS}
\usepackage{graphicx}
\usepackage{subfig}

\title{Search for Very-High-Energy (E $>$ 100 GeV) Emission from Geminga Supernova by VERITAS.}

\ShortTitle{Search for VHE Emission from Geminga Supernova by VERITAS.}

\author{\speaker{A. U. Abeysekara} for the VERITAS Collaboration\footnote{for collaboration list see PoS(ICRC2019)1177} \footnote{https://veritas.sao.arizona.edu}\\
        University of Utah\\
        E-mail: \email{a.abeysekara@utah.edu}}

\abstract{Geminga is a nearby (250 pc) middle-aged (spin-down time scale ~12,000 years) pulsar associated with a supernova remnant. Geminga has been a prime candidate for the origin of the unexpectedly high flux of cosmic-ray positrons above 10 GeV detected at Earth. Extended TeV gamma-ray emission from a 2-degree region around the Geminga pulsar was detected by the HAWC observatory, thus suggesting efficient, high-energy leptonic acceleration. \textit{Fermi}-LAT observations show that the density of GeV leptons in the TeV nebula is lower than predicted by single zone and two zone diffusion models constrained with the HAWC measurements. However, the energy gap between \textit{Fermi}-LAT and HAWC ($\sim$500 GeV to ~1 TeV) remains under-examined. The VERITAS gamma-ray observatory is sensitive in the energy range from 100 GeV to greater than 30 TeV, filling the gap between \textit{Fermi}-LAT and HAWC. Therefore, VERITAS measurements potentially provide missing information. VERITAS has observed Geminga for 93 hours since 2009 including 28 hours in the 2018/2019 season. However, the standard VERITAS data analysis techniques have insufficient sensitivity to sources extended at the scale of the HAWC detection, due to difficulties with background estimation. We developed the Matched Runs Method (MRM) for VERITAS analysis of spatially extended sources. MRM has been demonstrated to be an effective technique by applying it to archival VERITAS data, and we are currently applying it to the Geminga observations. Here we present the summary of the MRM.}

\FullConference{36th International Cosmic Ray Conference -ICRC2019-\\
		July 24th - August 1st, 2019\\
		Madison, WI, U.S.A.}

\begin{document}

\section{Introduction}

ATIC, PAMELA, \textit{Fermi}-LAT, and AMS, among others \cite{PAMELA PExcess, TS93 PExcess, HEAT94 PExcess, CAPRICE94 PExcess, AMS-01 PExcess, HEAT00 PExcess, Fermi PExcess, AMS-02 PExcess} have detected a much higher fraction of cosmic-ray positrons at high energies than accounted for by typical models of cosmic-ray production. 
Interpretations of this result as a signature of dark matter annihilation quickly fell into disfavor, replaced by scenarios in which a set of pulsars, or even a single nearby pulsar such as Geminga or Monogem \cite{SAS-II Detection,COS-B Detection}, could have produced the entire excess \cite{Yuksel,PExcess Theory}. 
At the same time, first Milagro and now HAWC have provided us with an ever-increasing sample of highly extended TeV $\gamma$-ray halos surrounding nearby pulsars \cite{Milagro Detection 2,2ndHAWCCatalog,LessonsFromHAWC}.

A recent paper from HAWC \cite{HAWCGeminga} uses the TeV halos around Geminga and Monogem to constrain the positron diffusion constant as a function of energy. From this the authors argued that these pulsars (either individually or in concert) cannot account for the positron excess at Earth. However, this argument depends on a few key assumptions. First, HAWC's study (performed in the 1-50 TeV energy range) primarily constrains the diffusion constant using $>10$ TeV lepton energies. The HAWC result also depends on their assumption of homogeneous and isotropic diffusion, an assumption criticized by Profumo et. al. \cite{LessonsFromHAWC}. In fact, the positron diffusion constant can vary radially, with positrons that have escaped the halo experiencing a higher diffusion constant. 

In such a picture, Geminga remains a viable source of the positron excess. Tests of inhomogeneous and anisotropic models are currently stymied by large uncertainties in HAWC's measurement of the surface brightness near the pulsar. With adequate exposure, VERITAS' superior angular and energy resolution would decrease these uncertainties. However, the standard VERITAS background estimation techniques have insufficient sensitivity to sources extended at the scale of the HAWC detection, due to difficulties with background estimation. This article discusses the Matched Runs Method (MRM) for VERITAS analysis of spatially extended sources, and a summary of the VERITAS observations on Geminga.

\section{VERITAS Standard Background Estimation Techniques}

The ring background method (RBM) and reflected-region method (REM) are widely used to estimate the background of point-like sources \cite{RBMandREM}. RBM uses a ring around the Region of Interest (ROI) to estimate the level of background contamination in the ROI. REM uses a set of background regions with offsets from the camera center equals to that of ROI. Both of these techniques require a dedicated region in the camera's field of view (FOV) to estimate the background.  However, when the source extent becomes comparable to the FOV, insufficient area remains for defining background regions. Therefore, RBM and REM are not effective background estimation techniques to observe extended sources.

The ON/OFF method is an alternative technique that has been successfully employed by other gamma-ray observatories to observe spatially extended sources. This method observes a region on the ROI (ON observation) and a region away from the ROI (OFF observation), and uses the entire FOV of the OFF observation to estimate the background. Usually, OFF observations are taken on the same day in a sky patch with no known gamma-ray sources (blank field), and that follows the same path in elevation and azimuth. The ON/OFF technique is capable of measuring sources with spatial extent comparable to the FOV of the camera. However, this technique requires twice as much observation time compared to the RBM or the REM. Therefore, it is not practical to utilize IACTs to observe fainter sources, such as Geminga, that require deep observations. MRM offers a better alternative. 

\section{Matched Runs Method}

Similar to ON/OFF method, MRM also uses an OFF region to estimate the background. However, unlike ON/OFF method, MRM uses a ring or wobble offset regions in the OFF FOV to estimate the background. Therefore,  the OFF region does not have to be a blank field. Instead, the OFF observation can be pointed at a point-like source. This is an advantage of the MRM: observations of a point-like source can be used for OFF observations.

A disadvantage of the MRM is that, unlike the standard ON/OFF observation, there is no dedicated OFF run that follows the same path in elevation and azimuth. Instead, MRM searches the VERITAS database for observations that match the ON observation. The algorithm that searches for matched runs is fully automated, and the algorithm uses four parameters; difference in elevation angle $\Delta El$, the difference in azimuthal angle $\Delta Az$, the time gap $\Delta T$ between ON and OFF runs, and the difference between the number of non-gamma like events in the ON and OFF runs, $\Delta N^{CR}$. The standard for a matched run used here is $\Delta Az \leq 10^\circ$, $\Delta El \leq 10^\circ$, $\Delta T \leq 360$ days, and the matched run finding algorithm minimizes the $\Delta N^{CR}$ to a list of ON and OFF runs, 

\begin{equation}
    \Delta N^{CR} = \sum_{\textrm{All Runs}} |N^{CR}_{ON} - N^{CR}_{OFF}|
\end{equation}
where $N^{CR}_{ON}$ and $N^{CR}_{OFF}$ are the number of non-gamma like events that did not pass the standard VERITAS selection for gamma-like events for ON and OFF runs, respectively.

\section{Validation Tests}

\begin{figure}[!tbp]
  \centering
  \subfloat[Segue 1 Sky map]{\includegraphics[width=0.4\textwidth]{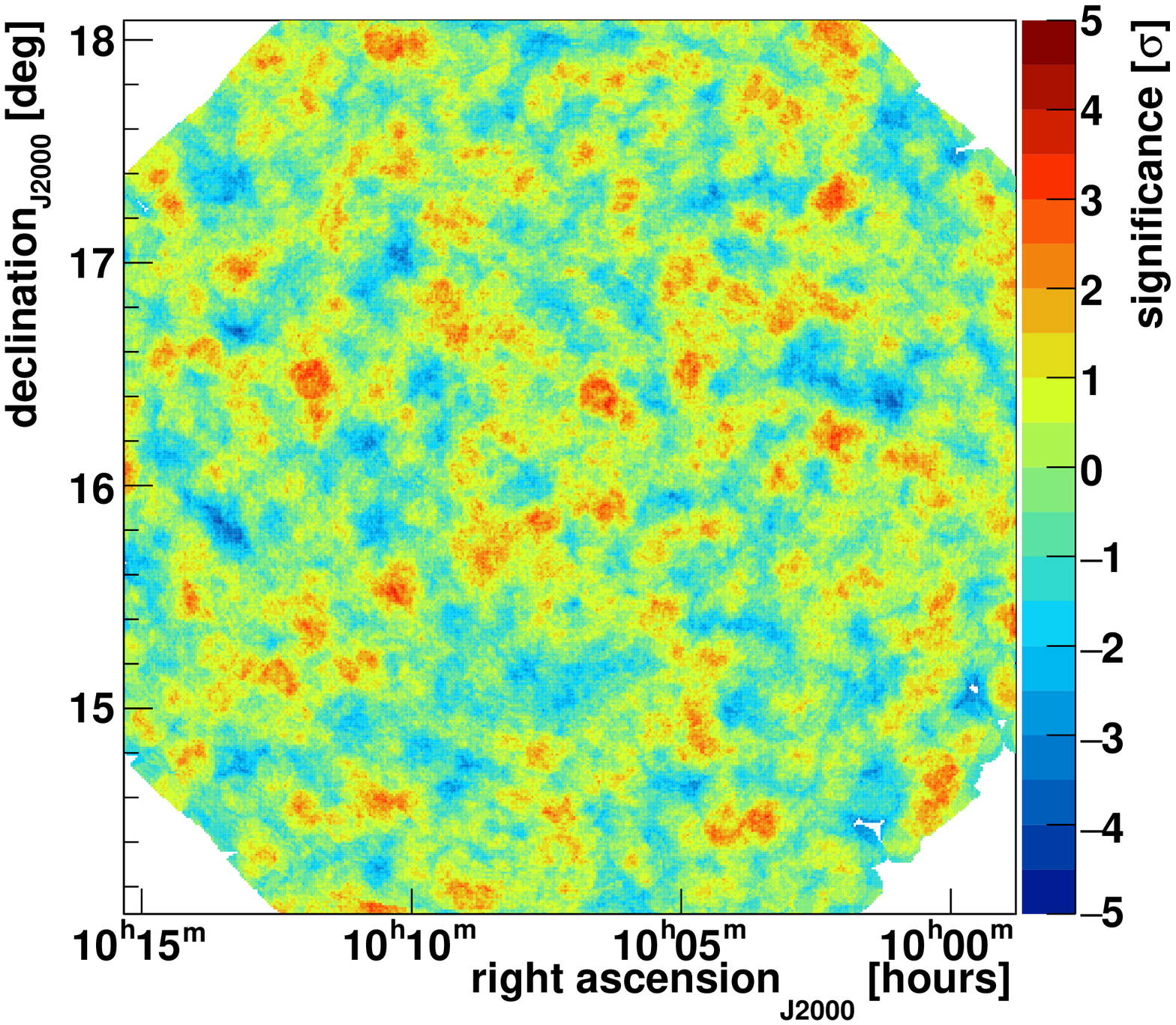}
  \label{SegueSkymapRE}}
  \hfill
  \subfloat[Segue 1 Significance distribution]{\includegraphics[width=0.4\textwidth]{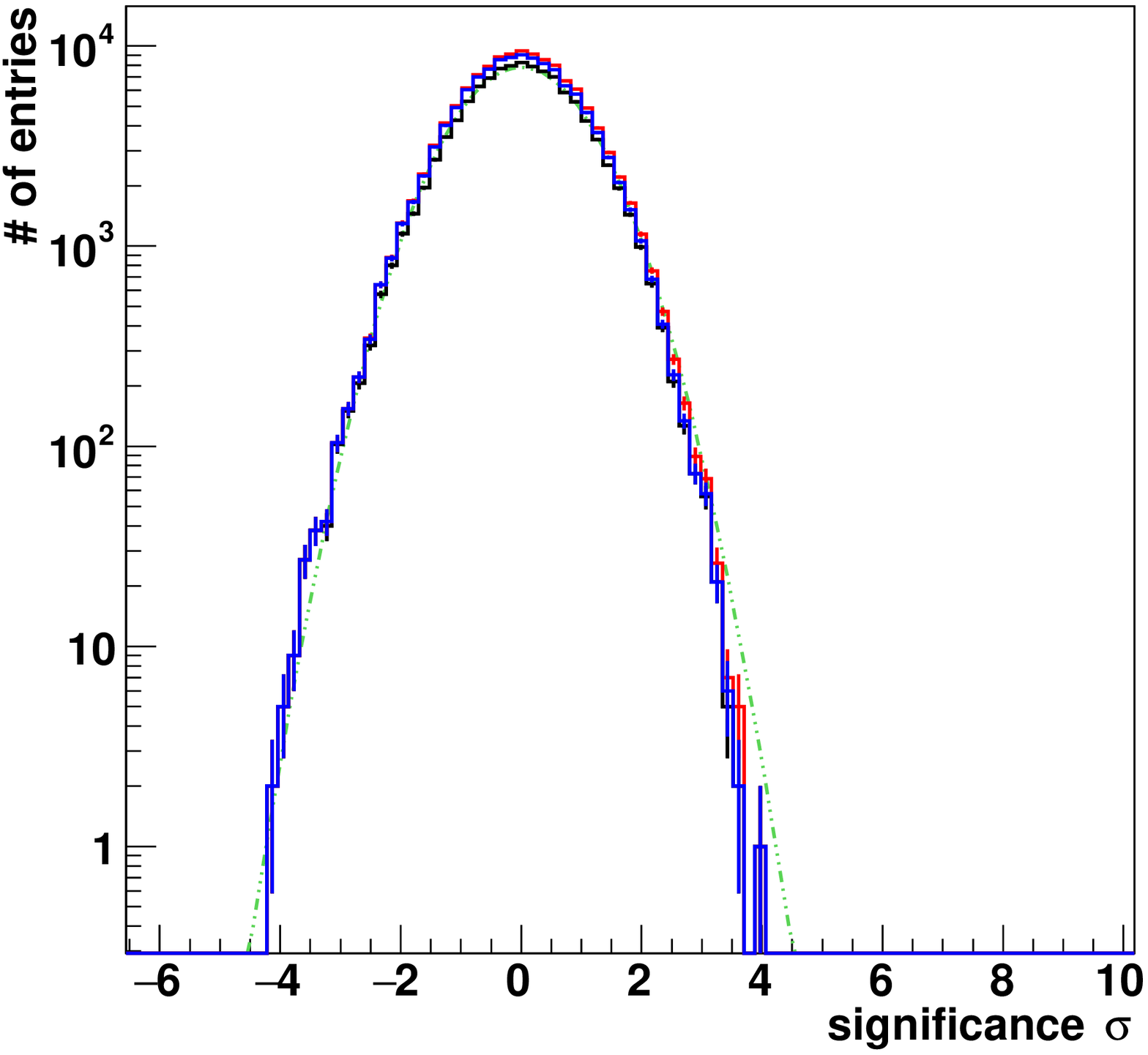}
  \label{SegueSignifRE}}
  \caption{Sky map and the significance distribution of the entire field of the view, around dSphs Segue 1, obtained with the standard background estimation technique REM.}
\end{figure}

During the past decade, VERITAS observed five dwarf spheroidal galaxies (dSphs) in order to search for dark matter \cite{VERITASDMProgram}. DSphs data has been analyzed with the standard VERITAS background estimation techniques, and no significant gamma-ray excess was observed in the direction of any of the five dSphs \cite{VERITASDMProgram}. Figure ~\ref{SegueSkymapRE} and Figure ~\ref{SegueSignifRE} shows the skymap and the significance distribution of the skymap obtained with twenty-six hours of Segue 1 observations analyzed with the standard REM background estimation technique. The significance distribution is consistent with a Gaussian distribution with a mean zero and standard deviation of one, which is expected when there is no gamma-ray source within the FOV, and excess events are due to background fluctuations.

\begin{figure}[!tbp]
  \centering
  \subfloat[Segue 1 Sky map]{\includegraphics[width=0.4\textwidth]{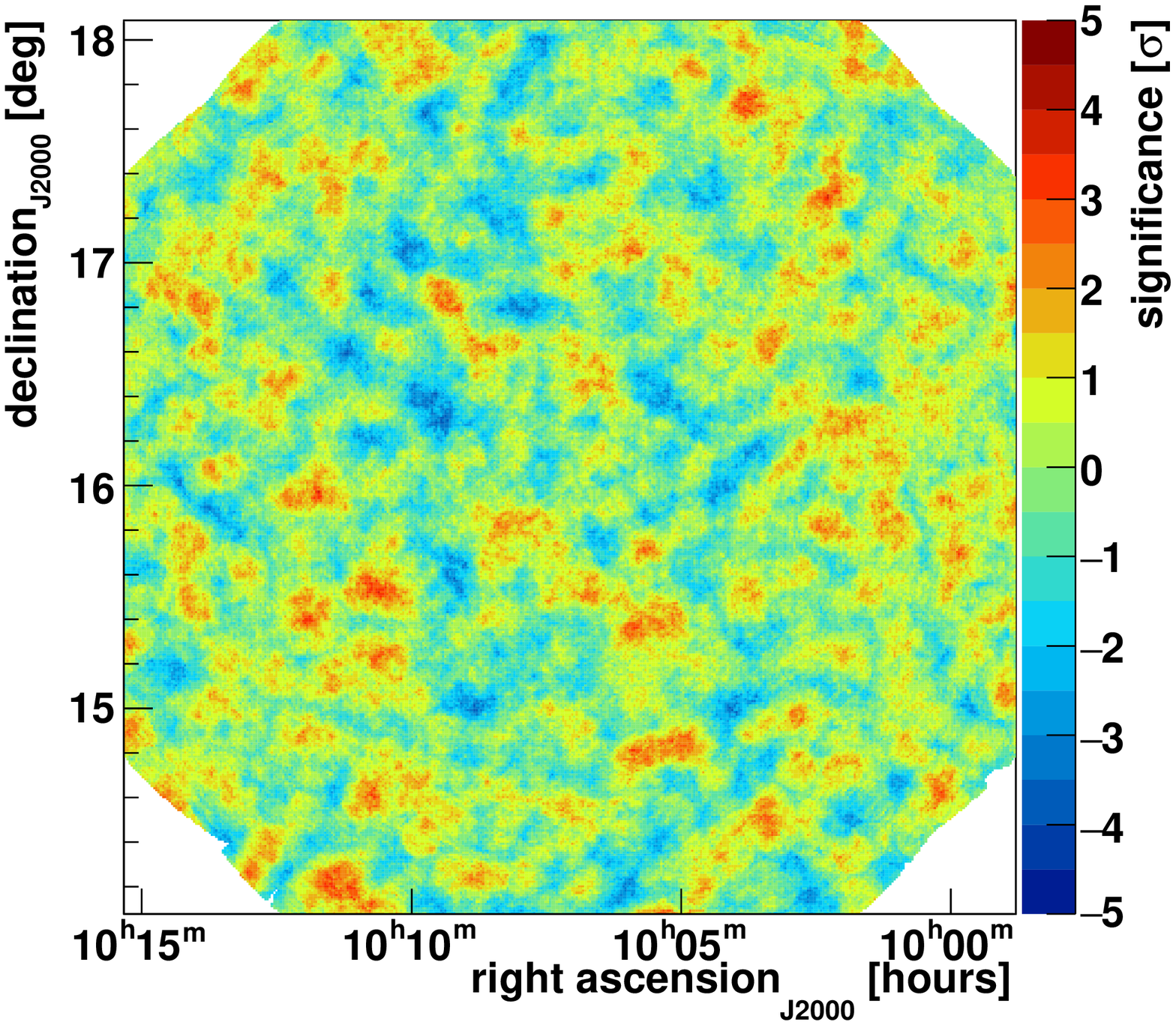}
  \label{SegueMRMSky}}
  \hfill
  \subfloat[Segue 1 Significance distribution]{\includegraphics[width=0.4\textwidth]{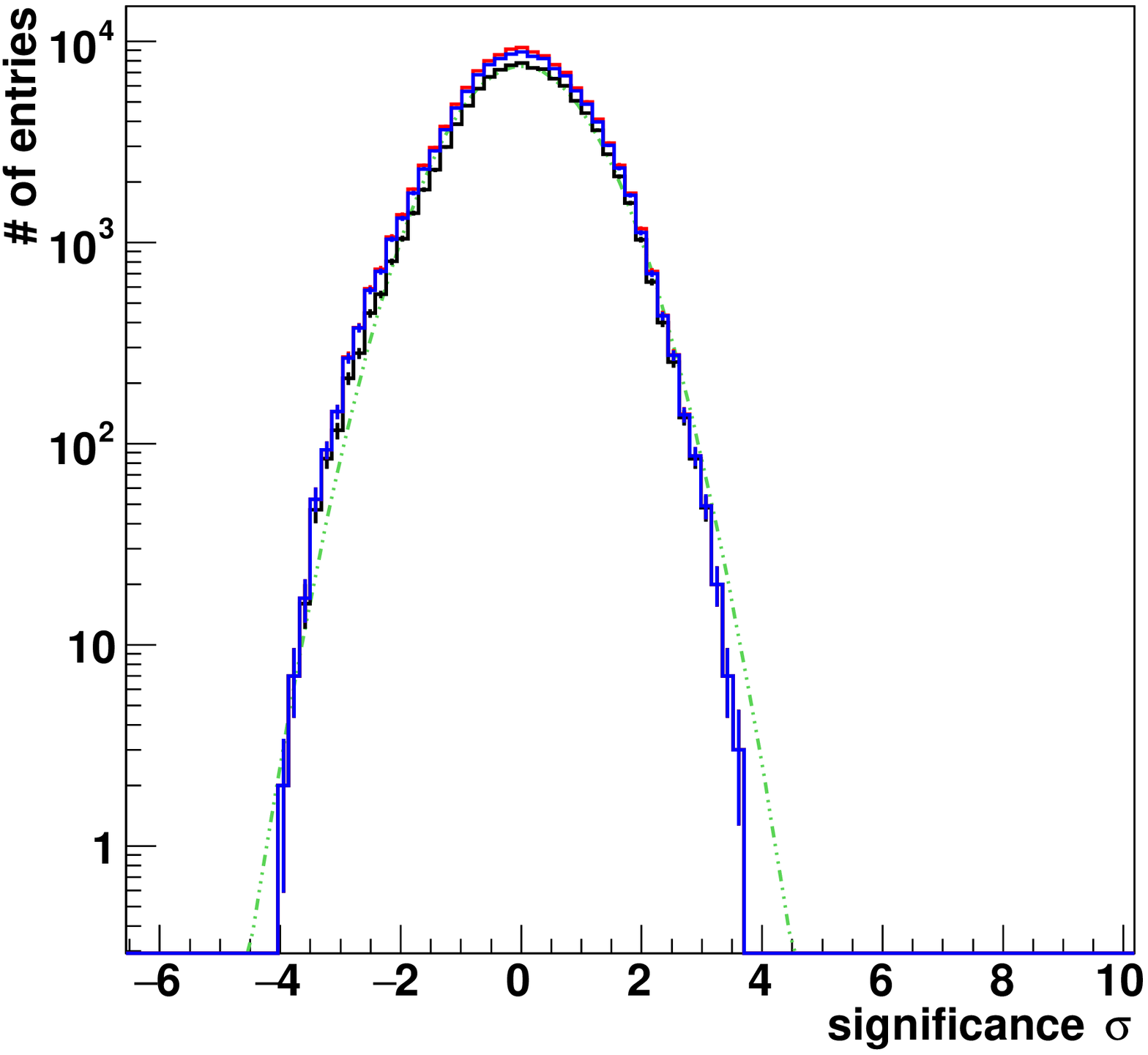}
  \label{SegueMRMSignif}}
  \caption{Sky map and the significance distribution of the entire field of the view, around dSphs Segue 1, obtained with MRM.}
\end{figure}

The Segue 1 data set has been analyzed with the MRM. Figure ~\ref{SegueMRMSky} and Figure ~\ref{SegueMRMSignif} show the sky map and the significance distribution of the FOV.  Matched runs for this analysis are coming from Crab and 1ES 0229 + 200 observations. The sky map obtained with the  MRM also does not show any features, and the significance distribution is consistent with a normal distribution ($\mu = 0 $, $\sigma = 1$).  Five hours of data of dSph Ursa Minor is also analyzed with MRM, and the significance distribution is consistent with a normal distribution. 

\section{Conclusion}
Validation tests with Segue 1 and Ursa Minor show that MRM is capable of estimating the background of the entire field-of-view. Currently, we are applying MRM to analyze weak blazars such as 1ES 0229+200, highly variable bright blazars such as Mrk 421, bright point-like sources such as Crab Pulsar Wind Nebula, and mildly extended sources such as IC 443. The MRM procedure will be applied to the Geminga data set when the validation tests are completed. 

\section{Acknowledgments}

This research is supported by grants from the U.S. Department of Energy Office of Science, the U.S. National Science Foundation and the Smithsonian Institution, and by NSERC in Canada. This research used resources provided by the Open Science Grid, which is supported by the National Science Foundation and the U.S. Department of Energy's Office of Science, and resources of the National Energy Research Scientific Computing Center (NERSC), a U.S. Department of Energy Office of Science User Facility operated under Contract No. DE-AC02-05CH11231. We acknowledge the excellent work of the technical support staff at the Fred Lawrence Whipple Observatory and at the collaborating institutions in the construction and operation of the instrument.

\end{document}